\documentclass[conference]{IEEEtran}
\usepackage{amssymb,amsmath,epsfig,graphicx,theorem}
\usepackage{amsfonts}
\usepackage{bm, color}
\newtheorem{Theo}{Theorem}
\newtheorem{Lem}{Lemma}

\newtheorem{Rem}{Remark}

\newtheorem{ex}{Example}
\usepackage{epsfig,rotating,setspace,latexsym,amsmath,epsf,amssymb,bm}
\usepackage{cite}
\usepackage{float}
\usepackage{subcaption}
\usepackage{cleveref}
\usepackage[linesnumbered,ruled,vlined]{algorithm2e}
\usepackage{stfloats}
\usepackage[numbers,sort&compress]{natbib}
\usepackage[T1]{fontenc}

\IEEEoverridecommandlockouts

\title{A Cyclic Placement Strategy for Multi-access Coded Caching}
\author{
\IEEEauthorblockN{Zeru Chen and Nan Liu}
  \IEEEauthorblockA{National Mobile Communications Research Laboratory\\
Southeast University\\
Nanjing, China 210096\\
Email: \{czr, nanliu\}@seu.edu.cn}
   \and
 \IEEEauthorblockN{Wei Kang}
   \IEEEauthorblockA{School of Information Science and Engineering\\
 Southeast University\\
 Nanjing, China 210096\\
 Email: wkang@seu.edu.cn}
 \thanks{
    This work is partially supported by the National Natural Science Foundation of China
    under Grants $62071115$ and $61971135$, and the Research Fund of National Mobile Communications Research Laboratory, 
    Southeast University (No. 2023A03).}}

\begin{document}

\maketitle

\begin{abstract}
    We investigate the multi-access coded caching problem, which involves $N$ files, $K$ users, and $K$ caches in this paper. 
    Each user can access $L$ adjacent caches in a cyclic manner. 
    We present a coded placement scheme for the case of cache $M=\frac{K-1}{KL}$, when $\frac{K-1}{L}$ is an integer. 
    The scheme is based on coded placement and involves cyclic placement in caches. 
    In many parameter settings, our scheme achieves a lower transmission rate compared to schemes without coded placement. 
    Additionally, the achieved transmission rate of the proposed scheme is optimal when $L=K-1$ and $N\leq K$. 
\end{abstract}

\section{Introduction}
With the rapid development of mobile communication and internet technologies, the data requirements for mobile internet devices is 
experiencing explosive growth. One approach to meet this demand is to prefetch and cache a portion of data on user devices, 
which can effectively utilize channel resources during non-peak hours and alleviate network load during peak periods. 
Maddah-Ali and Niesen introduced the concept of coded caching in their groundbreaking work \cite{Maddah:2014}. 
The proposed configuration for the content delivery network is outlined as follows: a central server has $N$ files, each with a unit size, 
and there are $K$ users, each equipped with a distinct cache having a capacity of $M$ units. During non-peak hours, the server fills a function of the 
file content into the caches without knowing the users' file demands. During peak hours, each user requests a file from the central server. 
The server, based on the requests and the content stored in the caches, broadcasts information to users through a shared 
error-free link. Our objective is to minimize the transmission rate of the server, while ensuring that each user can successfully recover 
their requested file.

\begin{figure}
    \includegraphics[width=0.9\linewidth]{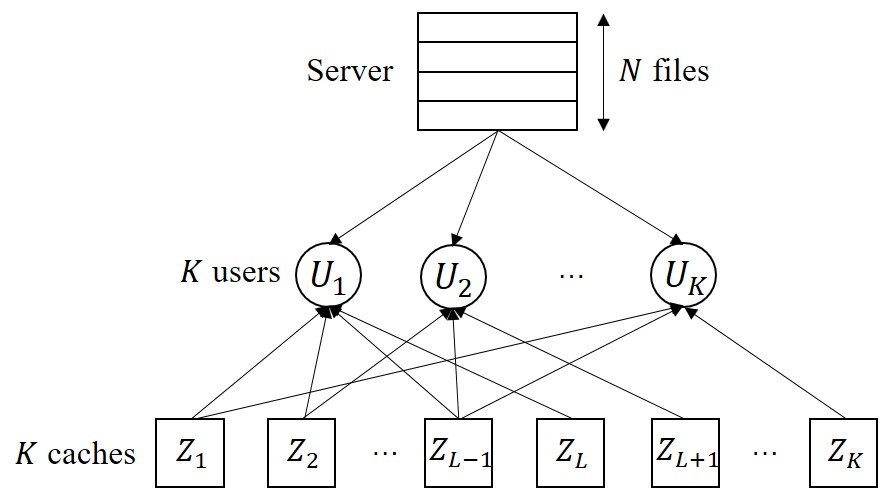}
    \caption{A $(N,K,L)$ multi-access coded caching system}
    \label{fig1}
\end{figure}

In practical scenarios, such as cellular networks, users can access multiple caches when their coverage areas overlap.
Driven by wireless networks, \cite{Hachem:2017} introduced the Multi-Access Coded Caching (MACC) problem. 
This problem involves a central server with $N$ files, $K$ caches, each having $M$ memory units, and $K$ users. 
Each user is connected to $L$ consecutive caches in a cyclic manner, as illustrated in Fig.\ref{fig1}. 
References \cite{Serbetci:2019,Reddy:2020,Cheng:2021,Mahesh:2021,Reddy:2021,sasi2021improved,Sasi:2021,Wang:2023,Reddy:2023,KP:2023} 
proposed various schemes for multi-access coded caching, but they are all limited to uncoded placement. 
The references \cite{Reddy:2020,Reddy:2021,Reddy:2023} transform the multi-access coded caching problem into the index coding problem.
Among them, \cite{Reddy:2020} proposed a coloring-based transmission scheme and obtained a converse bound for multi-access coded caching systems without coding 
placement when $L\geq K /2$ and $N\geq K$. Furthermore, when $L\geq K /2$ and $L=K-1$, $L=K-2$, $L=K-3$ ($K$ is even), 
or $L=K-K /s+1~(s\in \mathbb{N}^+)$, the proposed scheme achieves the optimal transmission rate $R^{*}$. 
References \cite{Reddy:2021,Reddy:2023} provided tighter upper bounds on the optimal rate-memory trade-off in the general case. 
Reference \cite{Serbetci:2019} investigated the case where $M=\frac{N(K-1)}{LK}$ and achieved the optimal transmission 
rate $R^{*}=\frac{1}{K}$. References \cite{Cheng:2021,Sasi:2021,Wang:2023} utilized Placement Delivery Arrays (PDAs) to construct 
some multi-access schemes. Reference \cite{KP:2023} introduced a scheme that achieves the optimal transmission rate 
$R^{*}=\frac{1}{K}$ under the condition of large cache, i.e., when 
$M\in\left[ \frac{N}{K-2}\cdot \frac{K-1}{K}, \frac{N}{K-2} \right]$. Reference \cite{Namboodiri2022coded} is the first paper to employ
 coded placement, and proposed a coded placement scheme for $M<(N-K+L)/K$, and this scheme 
was proven to be optimal when $N\leq K$.

In this paper, we investigate the multi-access coded caching problem and propose a cyclic placement-based coded placement scheme, which works when the cache size $M=\frac{K-1}{KL}$, where $\frac{K-1}{KL}$ is an integer. We show that the scheme
achieves a lower worst-case transmission rate than existing schemes when $N\leq K$. 

\emph{Notations}:

For a positive integer $n\in\mathbb{N}^+$, $[n]$ denotes the set $\{ 1,2,\dots,n \}$. 

$k$ and $K$ are two integers, then
$$
\left< k \right> _{K}=\begin{cases}
k~\mathrm{mod}~K, & \mathrm{if}~k~\mathrm{mod}~K\neq 0. \\
K, & \mathrm{if}~k~\mathrm{mod}~K=0.
\end{cases}
$$

For three integers $a$, $b$, and $K$, $[a:b]_{K}=\{\left< a \right>_{K},\left< a+1 \right>_{K},\dots,\left< b \right>_{K}\}$.

For a set $\mathcal{Z}$ and an index set $\mathcal{S}$, $|\mathcal{Z}|$ denotes the cardinality of $\mathcal{Z}$, and 
$\mathcal{Z}_{\mathcal{S}}$ denotes a set comprising the elements of $\mathcal{Z}$ positioned as specified by the index set $\mathcal{S}$.

The $\oplus$ represents the bitwise Exclusive OR (XOR) operation.

\section{System Model}
Let's consider the $(N,K,L)$ multi-access coded caching system, as shown in Fig. 1. A central server with $N$ files 
$\mathcal{W}=\{ W_{1},W_{2},\dots,W_{N} \}$, each having a unit size, is connected to $K$ users represented by 
$\mathcal{U}=\{ U_{1},U_{2},\dots,U_{K} \}$ through an error-free shared-link. The system has $K$ caches 
$\mathcal{Z}=\{ Z_{1},Z_{2},\dots,Z_{K} \}$, each with a memory size of $M$ units, and each user can access $L$ 
consecutive caches in a cyclic manner. Specifically, the user $U_{k},~k\in[K]$ has access to the caches 
$\mathcal{Z}_{k},\mathcal{Z}_{\left< k+1 \right>_{K}},\dots,\mathcal{Z}_{\left< k+L-1 \right>_{K}}$ .

During the placement phase, the server populates the caches with functions of file contents without prior knowledge of user demands. 
In the case of uncoded placement, each file is splitted into some subfiles, and specific subfiles are subsequently stored directly 
in each cache. In the coded placement scenario, functions of the file contents are stored in each cache. 
There are $K$ caching functions $\mu_{k}$, mapping the $N$ files into $Z_{k}$'s cache content, i.e., $Z_{k}\triangleq \mu_{k}(\mathcal{W}_{[N]})$, for $k\in[K]$.

In the delivery phase, each user requests a specific file from the server. Assuming that the user $U_{k}$ requests 
the file $W_{d(k)}$, where $k\in[K]$ and $d(k)\in[N]$, and the vector $\mathbf{d}=(d(1),d(2),\dots,d(K))$ denotes 
the user demands. After knowing the user demands, the server broadcasts messages $X_{\mathbf{d}}$ of total size $R_{\mathbf{d}}$ 
units to all users, where $X_{\mathbf{d}}$ is a function of $\mathcal{W}_{[N]}$, such that each user can recover the requested file. 
The delivery phase has $N^{K}$ encoding functions $\phi_{\mathbf{d}}$ and $KN^{K}$ decoding functions $\psi_{\mathbf{d},k}$. The encoding function determines the transmission message 
$X_{\mathbf{d}}$, i.e. $X_{\mathbf{d}}\triangleq \phi_{\mathbf{d}}(\mathcal{W}_{[N]})$, for $\mathbf{d}\in [N]^{K}$. 
And the decoding function helps the users to estimate their requested file, i.e., 
$\hat{W}_{d(k)}\triangleq \psi_{\mathbf{d},k}(X_{\mathbf{d}},\mathcal{Z}_{[k:k+L-1]_{K}})$, for $\mathbf{d}\in [N]^{K}$ and $k\in [K]$. 
The probability of error is defined as
$$
P_{e}=\underset{ \mathbf{d}\in[N]^{K} }{ \max }\underset{ k\in[K] }{ \max }P(\hat{W}_{d(k)}\neq W_{d(k)})
$$

The maximum transmission rate in the worst case, is defined as $R = \underset{\mathbf{d}\in[N]^{K}}{\max} R_{\mathbf{d}}$. 
Our goal is to minimize the maximum transmission rate $R$, ensuring that all users can recover their requested files without errors. 
The memory-rate pair $(M,R)$ is defined as achievable if there exists a scheme with $P_{e}<\varepsilon,~\forall \epsilon>0$, for any arbitrary users' demands. 
Correspondingly, the optimal rate-memory trade-off is defined as 
$R^{*}(M)=\mathrm{inf}\{ R:(M,R)\mathrm{~is~achievable} \}$.

\section{Main Results and Discussions}
In this section, we provide the main results of our multi-access coded caching scheme that leverage coded 
placement techniques. Then, we will compare the performance of our proposed scheme with existing results and provide some discussions.
\begin{Theo} \label{main}
    For a $(N,K,L)$ multi-access coded caching scheme with $N\leq K$, if $\frac{K-1}{L}$ is an integer, the memory-rate pair $\left( \frac{K-1}{KL},N-1 \right)$ 
    is achievable.
\end{Theo}

\begin{figure}
    \centering
    \begin{subfigure}{0.48\textwidth}
        \centering   
        \includegraphics[width=1\linewidth]{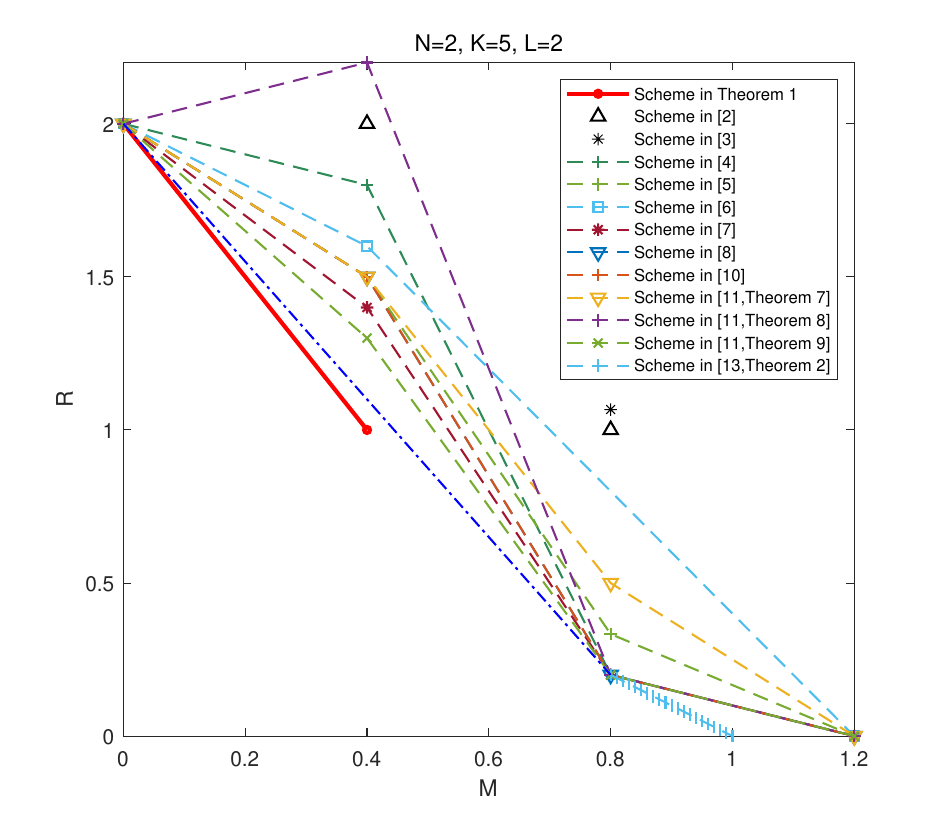}
        \caption{$(N=2,K=5,L=2)$}
        \label{fig2:sub1}
    \end{subfigure}
    
    \begin{subfigure}{0.48\textwidth}
        \centering   
        \includegraphics[width=1\linewidth]{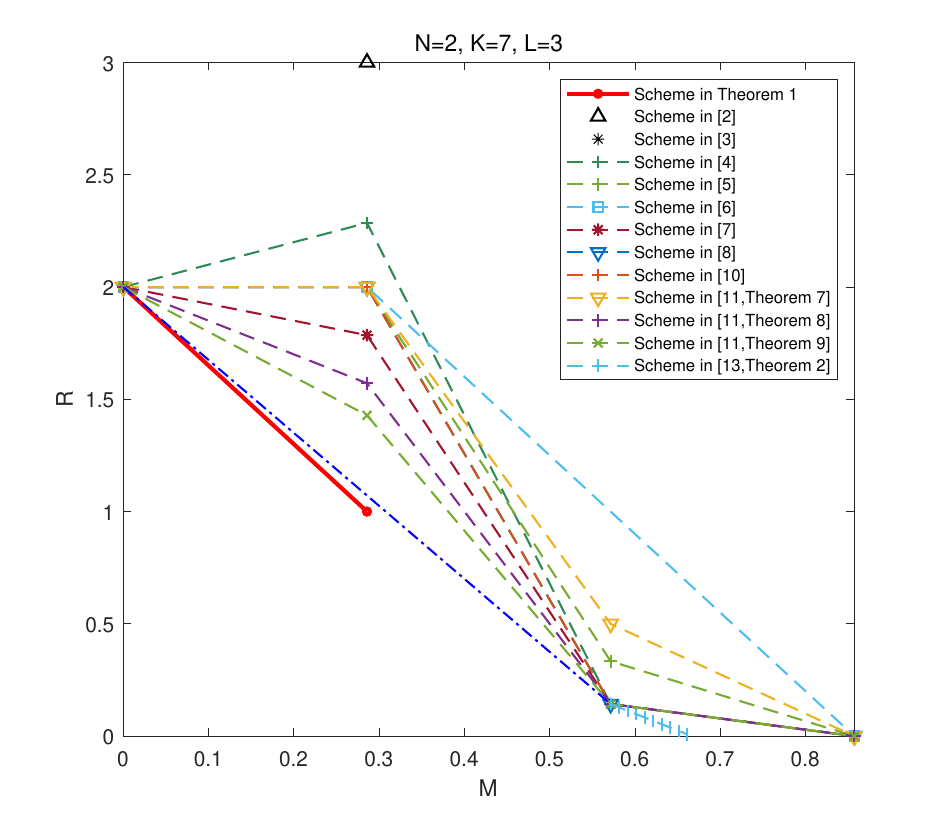}
        \caption{$(N=2,K=7,L=3)$}
        \label{fig2:sub2}
    \end{subfigure}

    \begin{subfigure}{0.48\textwidth}
        \centering   
        \includegraphics[width=1\linewidth]{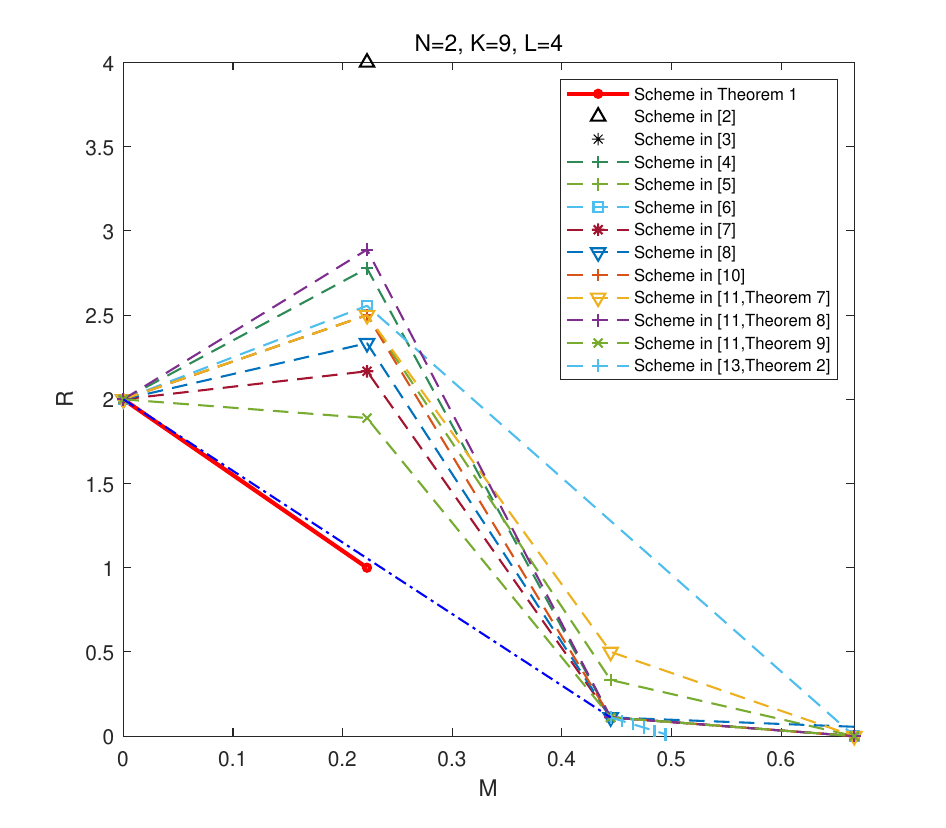}
            \caption{$(N=2,K=9,L=4)$}
            \label{fig2:sub3}
    \end{subfigure}
    
    \caption{
    \label{fig2:total}
    Comparison with existing research results
    }
\end{figure}

The proof of the theorem will be provided in Section IV. Various multi-access coded caching schemes from the existing 
literature are compared with our proposed schemes in Fig.\ref{fig2:total}. When $M=0$, $R=\min{(N,K)}$ is a trivial point 
which means that the users have no caches and the server need to transmit the files directly to satisfy the users' requests.

Since reference \cite{Hachem:2017,Serbetci:2019,Reddy:2020,Cheng:2021,Mahesh:2021,Reddy:2021,sasi2021improved,Wang:2023,Reddy:2023} 
did not specifically consider the case of $N<K$, there may be instances where some $(M,R)$ pairs 
can be achieved by these schemes, but is not on the convex envelope of the rate-memory tradeoff points. The dashed 
lines in Fig.\ref{fig2:total} represent the achievable $R(M)$ proposed in these references, while the dash dot line represents 
the lower convex envelope of the rate-memory trade-off points. Therefore, at the cache point $M=\frac{K-1}{KL}$, schemes in \cite{Hachem:2017,Serbetci:2019,Reddy:2020,Cheng:2021,Mahesh:2021,Reddy:2021,sasi2021improved,Wang:2023,Reddy:2023}
 can achieve a transmission rate point through memory sharing. In contrast, the coded placement scheme proposed in 
this paper, whose performance is denoted by the solid line, can achieve a lower transmission rates $R$ at the cache point $M=\frac{K-1}{KL}$ in 
certain special cases, such as those shown in Fig. \ref{fig2:total} where $(N=2,K=5,L=2)$, $(N=2,K=7,L=3)$, and $(N=2,K=9,L=4)$.

\begin{Lem}
For $N\leq K$, $L=K-1$, $R^{*}=N-1$ is the optimal rate at the memory point $M=\frac{1}{K}$.
\end{Lem}

\emph{Proof}:
    A converse for the multi-access coded caching problem is given in \cite[Theorem 2]{Namboodiri2022lower}, and it matches the achievability result of the proposed
 Theorem \ref{main} above. Hence, we may concluded that the $R^{*}(M)=N-1$ when $M=\frac{1}{K}$.
\begin{Rem}
    The rate-memory tradeoff point $(\frac{1}{K},N-1)$ in the $(N,K,L=K-1)$ multi-access problem where $N\leq K$ can also be achieved in 
    \cite{Namboodiri2022coded} with memory sharing. When $N=2$, the proposed scheme in Theorem \ref{main} is the same as the one in \cite{Namboodiri2022coded}. 
    When $N>2$, the proposed scheme is different from \cite{Namboodiri2022coded}. 
\end{Rem}

\section{Proof of Theorem \ref{main}}
Theorem \ref{main} is proved by presenting a multi-access coded caching scheme for the cache size $M=\frac{K-1}{KL}$, 
achieving the transmission rate $R=N-1$. Note that this scheme is valid only when $\frac{K-1}{L}$ is an integer.
\begin{itemize}
    \item \textbf{Placement Phase}: In the placement phase, the server splits the file $W_{n},~n\in[N]$ into $K$ non-overlapping 
    subfiles of equal size, i.e. $W_{n}=(W_{n,1},W_{n,2},\dots,W_{n,K})$. Let $F_{j}=\bigoplus_{n=1}^{N}W_{n,j},~j\in [K]$
     be the $j$-th coded file. That is, the term $F_{j}$ denotes the bitwise XOR sum of all subfiles $W_{n}$'s at position 
     $j$, $W_{n,j}$. Denote the set of coded files as $\mathcal{F}=\{ F_{1},F_{2},\dots,F_{K} \}$. Subsequently, we employ a cyclic 
     placement strategy to store these coded files $F_{j}$ in the caches. The $k$-th cache, whose content is denoted as 
     $\mathcal{Z}_{k}$, stores $\{F_{\left< k+iL \right>_{K}}|i\in [\frac{K-1}{L}-1]\cup\{0\}\}$, $k\in [K]$. 
     Each cache is stored with $\frac{K-1}{L}$ coded files, and each coded files has a size of $\frac{1}{K}$ units, this requires the memory size $M=\frac{K-1}{KL}$. 
     For instance, in the case of a $(N,K=5,L=2)$ multi-access scenario, cache $\mathcal{Z}_{4}$ stores $\{ F_{4},F_{1} \}$.
      Since each user can cyclically access $L$ caches, we define $\mathcal{C}_{k}$ as the set of files accessible to 
      user $k$, i.e., $\mathcal{C}_{k}=\mathcal{Z}_{[k:k+L-1]_{K}}=\mathcal{F}_{[k: k+K-2]_{K}}$, $k\in[K]$. 
      For example, in the $(N,K=5,L=2)$ multi-access scenario, user $U_{5}$ can access $\mathcal{Z}_5=\{F_5,F_2\}$ and $\mathcal{Z}_1=\{F_1,F_3\}$, and therefore, the coded file set 
      $\mathcal{C}_{5}=\{ F_{5},F_{1},F_{2},F_{3} \}$.

      For each user, $U_{k}$ does not have $F_{\left< k-1 \right>_{K}}$ in their accessible files but has access to other 
      encoded files. Because there are $K$ coded files, and $U_k$ has $K-1$ such files $\mathcal{F}_{[k: k+K-2]_{K}}$, the only one it does not have is $F_{\left< k+K-1 \right>_K}$. And
      $F_{\left< k+K-1 \right>_K}=F_{\left< k-1 \right>_K}$. For instance, in the $(N,K=5,L=2)$ multi-access scenario, user $U_{1}$ does not have $F_{5}$. 
      For each coded file $F_{j}$, where $j\in[K]$, the user unable to access it is $U_{\left< j+1 \right>_{K}}$, 
      and the users who can access it are $\mathcal{U}_{[j+2:j+K]_{K}}$. For example, in the $(N,K=5,L=2)$ multi-access 
      scenario, $F_{5}$ is inaccessible to $U_{1}$ but can be accessed by $U_{2}, U_{3}, U_{4}, U_{5}$.

      For completeness, we formally describe the placement phase in Algorithm \ref{alg1}.

    \item \textbf{Delivery Phase}: We assume that each user $U_{i}$ demands file $W_{d(i)}$, where $d(i)\in[N]$. For each coded file 
    $F_{j},~j\in[K]$, the server transmits $W_{d(\left< j+1 \right>_{K}),j}$, enabling user $U_{\left< j+1 \right>_{K}}$ 
    to obtain a subfile of its requested file $W_{d(\left< j+1 \right>_{K})}$. For other users 
    $\mathcal{U}_{[j+2: j+K]_{K}}$ who possess $F_{j}$, there exists a subset $\mathcal{S}\subseteq [j+2: j+K]_{K}$ 
    such that $d(i)=d(\left< j+1 \right>_{K}),~\forall i\in\mathcal{S}$, meaning that other users requesting file 
    $W_{d(\left< j+1 \right>_{K})}$ can also obtain the necessary subfile, $W_{d(\left< j+1 \right>_{K}),j}$. 

    At this point, users $\mathcal{U}_{[j+2: j+K]_{K}}$ can decode 
    $F_{j}'=F_{j}\oplus W_{d(\left< j+1 \right>_{K}),j}=\bigoplus_{n=1,n\neq d(\left< j+1 \right>_{K})}^{N}W_{n,j}$. 
    So, there exists a set $\mathcal{T}\subset[N]\backslash d(\left< j+1 \right>_{K})$, where $|\mathcal{T}|=N-2$. 
    For each element $t$ in the set $\mathcal{T}$, the server transmits $W_{t,j}$, and as a result, the subfiles 
    $W_{m,j},~m\in[N]\backslash d(\left< j+1 \right>_{K})$ from $N-1$ files in $F_{j}'$ can be decoded. 
    All users in $\mathcal{U}_{[j+2: j+K]_{K}\backslash \mathcal{S}}$ can obtain the requested subfile $W_{d(i),j}$, 
    where $i\in[j+2: j+K]_{K}\backslash \mathcal{S}$, and $d(i)\in[N]\backslash d(\left< j+1 \right>_{K})$. 
    That is to say, for each $F_j$, after transmitting $W_{d(\left< j+1 \right>_{K}),j}$, the server then arbitrarily 
    transmits $N-2$ subfiles which are XORed in $F_j$. There are totally $K$ such $F_j$'s, and each transmitted subfile has a size of $1/K$, 
    then $R=K(N-1)\cdot \frac 1K=N-1$.

    Therefore, in the decoding process for an encoded file $F_{j}$, each user $i$ can obtain the requested subfile 
    $W_{d(i),j},~i\in[K]$. So, by decoding all the $F_{j},~j\in[K]$, user $i$ can obtain $W_{d(i),j},~i\in[K],~j\in[K]$, 
    thereby recovering the file $W_{d(i)}$. 
    
    Also for completeness, the delivery phase is shown in Algorithm \ref{alg2}.
\end{itemize}

\begin{table*}[t]
    \centering
    \begin{tabular}{|c|c|c|c|c|}
        \hline
        $\mathbf{Z_1}$ & $\mathbf{Z_2}$ & $\mathbf{Z_3}$ & $\mathbf{Z_4}$ & $\mathbf{Z_5}$ \\
        \hline
        $W_{1,1}\oplus W_{2,1}$ & $W_{1,2}\oplus W_{2,2}$ & $W_{1,3}\oplus W_{2,3}$ & $W_{1,4}\oplus W_{2,4}$ 
        & $W_{1,5}\oplus W_{2,5}$ \\
        $W_{1,3}\oplus W_{2,3}$ & $W_{1,4}\oplus W_{2,4}$ & $W_{1,5}\oplus W_{2,5}$ & $W_{1,1}\oplus W_{2,1}$ 
        & $W_{1,2}\oplus W_{2,2}$ \\
        \hline
    \end{tabular}
    \caption{Cache contents of $(N=2,K=5,L=2)$}
    \label{T1}
\end{table*}

\begin{table*}[t]
    \centering
    \begin{tabular}{|c|c|}
        \hline
        $\mathbf{d}$ & $X_{\mathbf{d}}$ \\
        \hline
        $(1,2,1,2,2)$ & $(W_{2,1},W_{2,3},W_{1,2},W_{2,4},W_{1,5})$ \\
        \hline
        $(1,1,2,2,2)$ & $(W_{1,1},W_{2,3},W_{2,2},W_{2,4},W_{1,5})$ \\
        \hline
        $(1,2,2,2,2)$ & $(W_{2,1},W_{2,3},W_{2,2},W_{2,4},W_{1,5})$ \\
        \hline
    \end{tabular}
    \caption{Transmissions of $(N=2,K=5,L=2)$}
    \label{T2}
\end{table*}

\begin{table*}[t]
    \centering
    \begin{tabular}{|c|c|c|c|c|}
        \hline
        $\mathbf{Z_1}$ & $\mathbf{Z_2}$ & $\mathbf{Z_3}$ & $\mathbf{Z_4}$ & $\mathbf{Z_5}$ \\
        \hline
        $W_{1,1}\oplus W_{2,1}\oplus W_{3,1}$ 
        & $W_{1,2}\oplus W_{2,2}\oplus W_{3,2}$ 
        & $W_{1,3}\oplus W_{2,3}\oplus W_{3,3}$ 
        & $W_{1,4}\oplus W_{2,4}\oplus W_{3,4}$ 
        & $W_{1,5}\oplus W_{2,5}\oplus W_{3,5}$ \\
        $W_{1,3}\oplus W_{2,3}\oplus W_{3,3}$
        & $W_{1,4}\oplus W_{2,4}\oplus W_{3,4}$ 
        & $W_{1,5}\oplus W_{2,5}\oplus W_{3,5}$
        & $W_{1,1}\oplus W_{2,1}\oplus W_{3,1}$ 
        & $W_{1,2}\oplus W_{2,2}\oplus W_{3,2}$ \\
        \hline
    \end{tabular}
    \caption{Cache contents of $(N=3,K=5,L=2)$}
    \label{T3}
\end{table*}

\begin{algorithm}[htbp]
    \SetAlgoLined
    \KwIn{$\mathcal{W}$, $K$, $L$}
    \KwOut{$\mathcal{Z}$}
    \textbf{Initialize:} $\mathcal{Z}_{k}\gets \{\}$, $\mathcal{F}\gets\{\}$

    \For{$j\in [K]$}{
        $F_{j}=\bigoplus_{n=1}^{N}W_{n,j}$\;
        $\mathcal{F}\gets \mathcal{F}\cup F_{j}$\;
    }
    
    \For{$k\in[K]$}{
        \For{$i\in [\frac{K-1}{L}-1]\cup\{0\}$}{
            $\mathcal{Z}_{k}\gets \mathcal{Z}_{k}\cup F_{\left< k+iL\right>_{K}}$\;
        }
    }
    
    \caption{Placement Phase, $M=\frac{K-1}{KL}$}
    \label{alg1}
\end{algorithm}

\begin{algorithm}[htbp]
    \SetAlgoLined
    \KwIn{$N$, $K$, $\mathbf{d}=(d(1),d(2),\dots,d(K))$}
    \KwOut{$X_{\mathbf{d}}$}
    \textbf{Initialize:} $X_{\mathbf{d}}\gets\{\}$

    \For{$j\in [K]$}{
        $X_{\mathbf{d}}\gets X_{\mathbf{d}}\cup W_{d(\left< j+1 \right>_{K}),j}$\;
        $\exists \mathcal{T}\subset[N]\backslash d(\left< j+1 \right>_{K}),~|\mathcal{T}|=N-2$\;
        \For{$t\in \mathcal{T}$}{
            $X_{\mathbf{d}}\gets X_{\mathbf{d}}\cup W_{t,j}$\;
        }
    }
    \caption{Delivery Phase}
    \label{alg2}
\end{algorithm}

\section{Examples}
In this section, we provide two examples to illustrate in detail our placement and transmission strategies.
\begin{ex}
    Consider $(N=2,K=5,L=2)$ multi-access coded caching. Let $M=\frac{K-1}{KL}=\frac{2}{5}$. 
    In the placement phase, the server splits each file into 5 non-overlapping sub-files of equal size, i.e., 
    $W_{n}=(W_{n,1},W_{n,2},W_{n,3},W_{n,4},W_{n,5})$ for all $n\in [5]$. Then $F_{j}=W_{1,j}\oplus W_{2,j}$, 
    for $j\in[K]$. Then the server fills the caches as in Table \ref{T1}, by the rule 
    $\mathcal{Z}_{k}=\{ F_{\left< k+iL \right>_{K}}:i\in [\frac{K-1}{L}-1]\cup{0} \},~k\in[K]$.

    Let $\mathbf{d}=(d_{1},d_{2},d_{3},d_{4},d_{5})$ be the demands vector. Consider the case 
    $\mathbf{d}=(1,2,1,2,2)$. For the coded file $F_{2}=W_{1,2}\oplus W_{2,2}$, which is not accessible to user $U_{3}$. 
    Therefore, the server transmitts $W_{1,2}$ so that $U_{3}$ can get its requested subfile. Meanwhile, 
    $U_{1}$ gets its requested subfile as well. On the other hand, $U_{2}$, $U_{4}$, $U_{5}$ can decode $W_{2,2}$, 
    which is the required subfile of these three users.
    Table \ref{T2} gives the transmission 
    $X_{\mathbf{d}}$ under different user demands, and other cases can be easily getted by using symmetry.
\end{ex}

\begin{ex}
    Consider $(N=3,K=5,L=2)$ multi-access coded caching. Let $M=\frac{K-1}{KL}=\frac{2}{5}$. 
    In the placement phase, the server splits each file into 5 non-overlapping sub-files of equal size, i.e., 
    $W_{n}=(W_{n,1},W_{n,2},W_{n,3},W_{n,4},W_{n,5})$ for all $n\in [5]$. 
    Then $F_{j}=W_{1,j}\oplus W_{2,j}\oplus W_{3,j}$,  for $j\in[K]$. Then the server fills the caches as in Table \ref{T3}, 
    by the rule $\mathcal{Z}_{k}=\{ F_{\left< k+iL \right>_{K}}:i\in [\frac{K-1}{L}-1]\cup{0} \},~k\in[K]$.

    Assume that the demands vector $\mathbf{d}=(1,2,3,1,2)$. Consider the coded file 
    $F_{1}=W_{1,1}\oplus W_{2,1}\oplus W_{3,1}$, which is not accessible to user $U_{2}$. 
    Therefore, the server transmitts $W_{2,1}$ so that $U_{2}$ can get its requested subfile. Meanwhile, 
    $U_{5}$ gets its requested subfile as well. Then users $U_{1}$, $U_{3}$, $U_{4}$ have $F_{1}'=W_{1,1}\oplus W_{3,1}$, 
    and the server transmitting either $W_{1,1}$ or $W_{3,1}$ can make these users get their requested subfiles. 
    For other coded files $F_ {j}$, the server's transmission is as the same. For the case $\mathbf{d}=(1,2,3,1,2)$, 
    the server transmits $X_{\mathbf{d}}=(W_{2,1},W_{1,1},W_{1,3},W_{2,3},W_{3,2},W_{2,2},W_{2,4},W_{1,4},W_{1,5},W_{2,5})$ 
    to satisfy all the users' needs.
\end{ex}


\section{Conclusions}
In this paper, we investigated the multi-access coded caching problem. We employed a coded placement scheme that, 
in many parameter settings, achieves a lower transmission rate compared to schemes without coded placement. 
Additionally, the achieved transmission rate of the proposed scheme is optimal when $L=K-1$ and $N\leq K$.

\bibliographystyle{unsrt}
\bibliography{ref}
\end{document}